\documentclass[
    ,final            
    ,sort&compress    
  ]
  {aipproc}

\layoutstyle{6x9}

\begin{document}

\newcommand{\msun}{M_\odot}
\newcommand{\mbh}{M_\bullet}
\newcommand{\mbul}{M_{\mathrm{bulge}}}
\newcommand{\hst}{\textit{HST\/}}
\newcommand{\inv}{$^{-1}$}
\newcommand{\cmsq}{cm$^{-2}$}
\newcommand{\ecsa}{erg cm$^{-2}$ s$^{-1}$ \AA$^{-1}$}
\newcommand{\ecsh}{erg cm$^{-2}$ s$^{-1}$ Hz$^{-1}$}
\newcommand{\ecs}{erg cm$^{-2}$ s$^{-1}$}
\newcommand{\es}{erg s$^{-1}$}
\newcommand{\cxo}{\textit{Chandra\/}}

\newcommand{\apj}{ApJ}
\newcommand{\apjs}{ApJS}
\newcommand{\apjl}{ApJ}
\newcommand{\mnras}{MNRAS}
\newcommand{\pasp}{PASP}
\newcommand{\aap}{A\&A}
\newcommand{\aj}{AJ}

\title{Detecting Low-Mass Supermassive Black Holes}


\author{Himel Ghosh}{
  address={Department of Astronomy, The Ohio State University, 140 W
  18th Ave, Columbus, OH 43210, USA}
}

\author{Smita Mathur}{
  address={Department of Astronomy, The Ohio State University, 140 W
  18th Ave, Columbus, OH 43210, USA}
}

\author{Fabrizio Fiore}{
  address={INAF - Osservatorio Astronomico di Roma, via Frascati 33, 00040
Monteporzio Catone (Roma), Italy}
}

\author{Laura Ferrarese}{
  address={Herzberg Institute of Astrophysics, 5071 West Saanich Road, Victoria, BC V8X 4M6, Canada}
}

\begin{abstract}
We demonstrate the feasibility of uncovering supermassive black holes
in late-type, ``quiescent'' spiral galaxies by detecting signs
of very low-level nuclear activity. We use a combination of x-ray
selection and multi-wavelength follow-up. Here, we apply this
technique to NGC 3184 and NGC 5457, both of type Scd, and show that
strong arguments can be made that both host AGNs. 
\end{abstract}

\maketitle


\section{Introduction}

 As discussed in Mathur et al.\ (these Proceedings) and in \cite{gmff08},
 the only way to detect low-mass supermassive black holes (SMBHs) may
 be by their accretion activity. We choose to use x-ray selection to
 identify candidate AGNs for the following reasons: First, x-ray
 emission is a hallmark of AGNs. Second, x-rays can penetrate
 obscuring material which may be hiding the line emitting
 regions. Third, there are fewer sources of x-rays in a galaxy than
 there are of optical and UV emission and so dilution of the AGN
 signature by host galaxy light is less of a problem, even if the AGN
 is moderately obscured.
Fourth, even if, as expected
in some theories \citep{n00,l03} AGNs that have
luminosities or accretion rates below a cut-off value do not have
broad-line regions, they should still be detectable in x-rays.
The disadvantage is that x-ray
observations by themselves cannot always distinguish between AGNs and
other x-ray sources, such as x-ray binaries (XRBs) and ultraluminous
x-ray sources (ULXs). Multi-wavelength data are needed to determine
the type of source. 

As examples, we present here two late-type spiral galaxies, NGC 3184
and NGC 5457 (M101). Their nuclear optical spectra show no signs of
AGNs \cite{hfs97-3}. But using archival \cxo\ data and
multi-wavelength data from the literature, we show that the sum total
of evidence strongly suggests that the galaxies host AGNs.

\section{NGC 3184}

\begin{figure}
  \includegraphics[height=0.2\textheight]{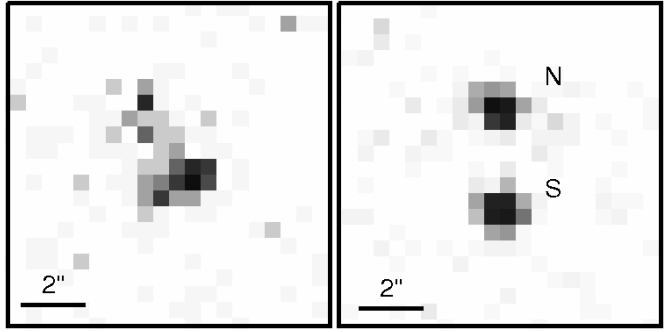}
\caption{\cxo\ images of the nuclei of NGC 3184 (left) and NGC 5457
  (right). Both panels are $10''$ on a side and have north up and east
  to the left. \textit{NGC 3184\/}: The nucleus consists of a point-like
  source and an extended component to the south of it. The bar on the
  lower left corresponds to a distance of $\sim\!80$ pc. \textit{NGC
  5457\/}: The nucleus is resolved into two sources, marked ``N'' and
  ``S'' in the figure. Source N is the active nucleus and source S is
  a star cluster. The bar on the lower left corresponds to a distance
  of $\sim\! 70$ pc.}
\label{fig:n31n54}
\end{figure}

\begin{figure}
  \includegraphics[height=.3\textheight]{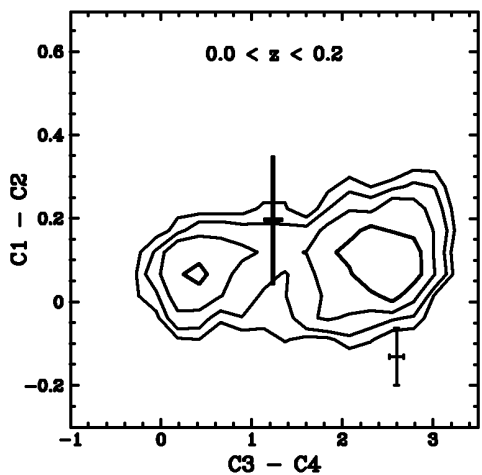}
  \caption{The unusually red mid-infrared color of the nucleus of NGC
  3184 is marked on a Spitzer IRAC color-color diagram. The contours
  show the distribution of $z\sim 0$ galaxies, separated by factors of
  1.8 in galaxy density. C1...C4 stand for the 3.6 $\mu$m, 4.5 $\mu$m,
  5.8 $\mu$m, 8.0 $\mu$m bands respectively. The left- and right-hand
  concentrations are spiral and elliptical galaxies respectively.  A
  typical error bar is shown in the lower right. (R.~J.~Assef 2007,
  priv. comm. Also see \cite{aea08}). \textit{Reprinted with
  permission of the author.}}
\label{fig:n3184}
\end{figure}

This galaxy is of type Scd, classified as having an \textsc{Hii}
nucleus by \citet{hfs97-3}, and is at a distance of 8.7 Mpc
\citep{t88}. The argument for the presence of an AGN is based on the
following: (a) \emph{The nucleus is an x-ray source}. Observations
with \cxo\ totaling 60 ks detect a point-like source with 36 counts
and a more extended component with 117 counts
(Fig.~\ref{fig:n31n54}). Assuming a power-law spectrum with slope
$\Gamma \sim 2$ and correcting for Galactic absorption only, the point
source has luminosity $L(0.3-8\mathrm{keV}) \sim 2\times 10^{37}$
\es. Luminosities in this range are typically associated with XRBs
rather than AGNs, but does not rule out an AGN, especially since
intrinsic absorption is unknown. (b) \emph{Infrared line ratios are
not that of an \textsc{Hii} nucleus but instead suggest the presence
of an AGN component}. A stronger argument that the source is an AGN
derives from infrared data. This galaxy is part of the
\textit{Spitzer} Infrared Nearby Galaxies Survey
\citep{kea03}. \citet{dea06} have used the equivalent width of the PAH
feature at $6.2 \mu$m and the fluxes in a mix of high- and
low-ionization IR lines
of S, Ne, and O to create diagnostic diagrams that
distinguish between AGN and star-forming galaxies. This nucleus falls
into the ``transition'' region between AGNs and \textsc{Hii}
regions. This suggests that the emission has an AGN component that
was diluted because of the large aperture used ($\sim 20''$)
to extract the fluxes. (c) \emph{Mid-infrared colors are redder than
those of normal galaxies}. The observed IRAC colors, $[3.6]-[4.5] =
+0.20$ $\pm 0.16$ and $[5.8]-[8.0] = +1.24$ $\pm 0.16$ (magnitudes in
Vega system) (D.~A.~Dale, priv.\ comm.), are redder than more than
80\% of normal late-type galaxies (Fig.~\ref{fig:n3184}; also see
\cite{aea08}). This is expected if there is an AGN, as AGN power-law
emission falls off more slowly than galactic emission in the
MIR. Thus, the IR line ratios and IRAC colors strongly argue that the
source is an AGN.

\section{NGC 5457}

\begin{figure}
\includegraphics[angle=-90,scale=0.75]{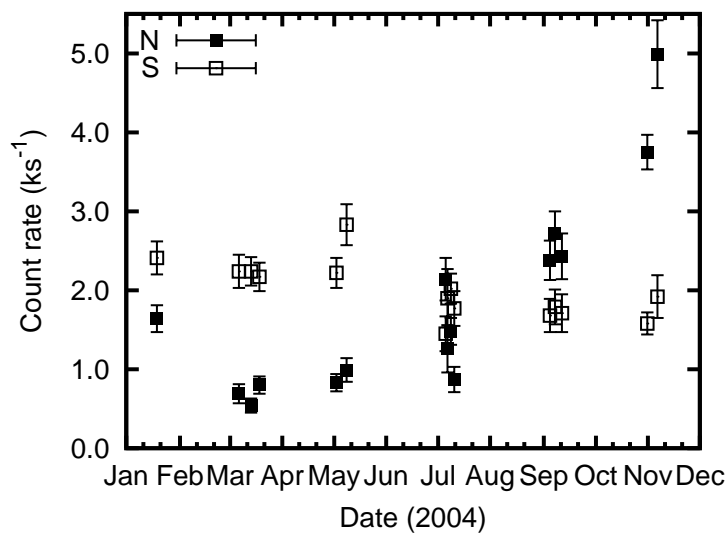}
\caption{X-ray variability of the nucleus of NGC 5457 as seen by
\cxo. The filled squares show the nuclear count rate, and the open
squares the count rate from a nearby star cluster. The nucleus
varies by almost a factor of ten between March and November, 2004.}
\label{fig:n5457}
\end{figure}

NGC 5457 is a galaxy of type Scd at a distance of approximately 7 Mpc
\citep{ssea98}. The nucleus was classified as \textsc{Hii} by
\citet{hfs97-3}. NGC 5457 was observed multiple times by \cxo\ for a
total observing time of approximately 1 Ms
(Fig.~\ref{fig:n31n54}). For this galaxy the x-ray data by themselves
provide evidence suggesting the presence of an AGN. (a) \emph{The
nucleus is an x-ray source}. (b) Even more importantly, \emph{the
nuclear source is variable}. The nucleus varies by about a factor of
10 over nine months (Fig.~\ref{fig:n5457}), with luminosity ranging
from $L(0.3-8\mathrm{keV}) \sim 3 \times 10^{37}$ to $3\times 10^{38}$
\es. The variability rules out multiple XRBs as the origin of the
x-ray emission, as they would have to be varying in phase. It is
possible for a single XRB to produce the observed x-ray emission, but
XRBs with luminosities this high are uncommon. (c) \emph{The x-ray
spectrum} shows an absorbed power-law with slope $\Gamma \approx 2$,
similar to what is seen in AGNs. The best-fit intrinsic absorption is
consistent with zero when the source is faint. (d) \emph{The ratios of
0.3--2 keV, 2--5 keV, and 5--8 keV counts} for this nucleus are
similar to those seen in Compton-thick AGNs \citep{lea06}. Points (c)
and (d) together suggest a highly obscured AGN where, like in the
Seyfert galaxy NGC 1068, the dominant observed emission is radiation
that has been scattered into our line of sight. Finally, (e) as can be
seen in Fig.~\ref{fig:n5457}, \emph{the behavior of the nucleus is
markedly different from that of a known star cluster} of comparable
x-ray luminosity which is $\sim\! 110$ pc away from the nucleus. This
suggests the nuclear source is not simply another star cluster.

\section{Conclusion}

There are compelling, though not conclusive, arguments that NGC 3184
and NGC 5457 actually host AGNs even though neither shows signs of
AGNs in their optical spectra. These AGNs are excellent candidates for
being low-mass SMBHs, as they reside in low-mass bulges. A similar
approach can be used to uncover more candidate low-mass SMBHs.


\begin{theacknowledgments}
We are grateful to D.~A.~Dale for kindly providing \textit{Spitzer}
fluxes for the nucleus of NGC 3184 prior to publication. Support for
this work was provided by the National Aeronautics and Space
Administration through Chandra Award Number GO7-8111X issued by the
Chandra X-ray Observatory Center, which is operated by the Smithsonian
Astrophysical Observatory for and on behalf of the National
Aeronautics Space Administration under contract NAS8-03060.

\end{theacknowledgments}



\bibliographystyle{aipproc}   


\end{document}